\documentclass[12pt]{article}
\usepackage{graphicx}
\usepackage{xspace}

\usepackage[colorlinks=true, urlcolor=blue, citecolor=black]{hyperref}  

\newcommand\pubnumber{DPF2015-60}
\newcommand\pubdate{\today}

\newcommand{\MET}{\ensuremath{E_{\mathrm{T}}^{\rm{miss}}}\xspace}

\newcommand{\pt}{\ensuremath{p_{\rm T}}\xspace}
\newcommand{\Ecalo}{\ensuremath{E_{\rm calo}}\xspace}

\def\osu{Department of Physics \\
The Ohio State University, Columbus, Ohio 43210}
\def\support{\footnote{On behalf of the CMS Collaboration}}

\def\Title#1{\begin{center} {\Large #1 } \end{center}}
\def\Author#1{\begin{center}{ \sc #1} \end{center}}
\def\Address#1{\begin{center}{ \it #1} \end{center}}

\newcommand\pubblock{\rightline{\begin{tabular}{l} \pubnumber\\
         \pubdate  \end{tabular}}}
\newenvironment{Abstract}{\begin{quotation}  }{\end{quotation}}
\newenvironment{Presented}{\begin{quotation} \begin{center} 
             PRESENTED AT\end{center}\bigskip 
      \begin{center}\begin{large}}{\end{large}\end{center} \end{quotation}}
\def\Acknowledgments{\bigskip  \bigskip \begin{center} \begin{large}
             \bf ACKNOWLEDGMENTS \end{large}\end{center}}

\def\beq{\begin{equation}}
\def\eeq#1{\label{#1}\end{equation}}
\def\eeqn{\end{equation}}

\def\beqa{\begin{eqnarray}}
\def\eeqa#1{\label{#1}\end{eqnarray}}
\def\eeqan{\end{eqnarray}}

\let\bar=\overbar

\def\Dslash{\not{\hbox{\kern-4pt $D$}}}
\def\dslash{\not{\hbox{\kern-2pt $\del$}}}

\def\msb{{\bar{\ssstyle M \kern -1pt S}}}

\begin{document}
\begin{titlepage}
\pubblock

\vfill
\Title{Search for disappearing tracks} 
\vfill
\Author{ H. Wells Wulsin \support}
\Address{\osu}
\vfill
\begin{Abstract}
We present a search for long-lived charged particles that decay within the CMS detector and produce the signature of a disappearing track. Disappearing tracks are identified as those with little or no associated calorimeter energy deposits and with missing hits in the outer layers of the tracker. The search uses proton-proton collision data recorded at $\sqrt{s} = 8$ TeV that corresponds to an integrated luminosity of 19.5 fb$^{-1}$. The results of the search are interpreted in the context of the anomaly-mediated supersymmetry breaking model and in terms of the phenomenological MSSM.
\end{Abstract}
\vfill
\begin{Presented}
DPF 2015\\
The Meeting of the American Physical Society\\
Division of Particles and Fields\\
Ann Arbor, Michigan, August 4--8, 2015\\
\end{Presented}
\vfill
\end{titlepage}
\def\thefootnote{\fnsymbol{footnote}}
\setcounter{footnote}{0}

\section{Introduction}
Despite a broad search program at the LHC, no clear signs of physics beyond the Standard Model have yet been uncovered.
However, exotic new particles may have evaded many of the searches to date if they have a significant lifetime, such that they travel a measurable distance before decaying.  Most LHC searches for supersymmetry or exotic particles apply stringent impact parameter requirements on the physics objects of interest, and thus may not be sensitive to long-lived particles.  

In this talk we present a search for long-lived charged particles that produce the signature of a disappearing track~\cite{EXO12034}.
The search is performed with a sample of proton-proton collisions recorded at $\sqrt{s} = 8$ TeV by the CMS detector~\cite{CMSDet} at the LHC, corresponding to an integrated luminosity of 19.5 fb$^{-1}$.    
A disappearing track could be produced by an exotic particle with decay products that are undetected because they are only weakly interacting or have very low momentum.  The observation of disappearing tracks would be a compelling sign of new physics, since there are few Standard Model processes that could produce such a signature.  If a signal were observed, there would be multiple ways to characterize the new particle, including its lifetime, its mass, and its decay products.  

To assess the search sensitivity we use a benchmark model, Anomaly Mediated Supersymmetry Breaking (AMSB).   In this model, the lightest chargino and neutralino are nearly degenerate, leading to a long lifetime of the chargino, on the order of $\sim$nanoseconds.  The chargino decays to the neutralino, which is undetected because it is weakly interacting, and a low-momentum charged pion, which is undetected because its curvature radius is too small to be reconstructed by the standard tracking algorithms.  Additionally, there are other models that predict particles that could be identified as disappearing tracks.  

\section{Event selection}  
Events are recorded with triggers with various thresholds on \MET (95--120 GeV), some of which additionally require a central jet with $\pt > 80$ GeV.  Signal events produce \MET when the exotic particles recoil from an ISR jet.  Offline we require events to have $\MET > 100$ GeV.  Several jet cleaning criteria are applied to reject QCD multijets backgrounds and events with spurious \MET.  Next we select a candidate track that is high-quality, high-momentum ($\pt > 50$ GeV), central ($|\eta| < 2.1$), and isolated from other tracks.  It must not be associated with any reconstructed electron, muon, or tau lepton.  

\begin{figure}[!h]
\centering
\begin{minipage}{0.45\linewidth}
\centerline{\includegraphics[width=\linewidth]{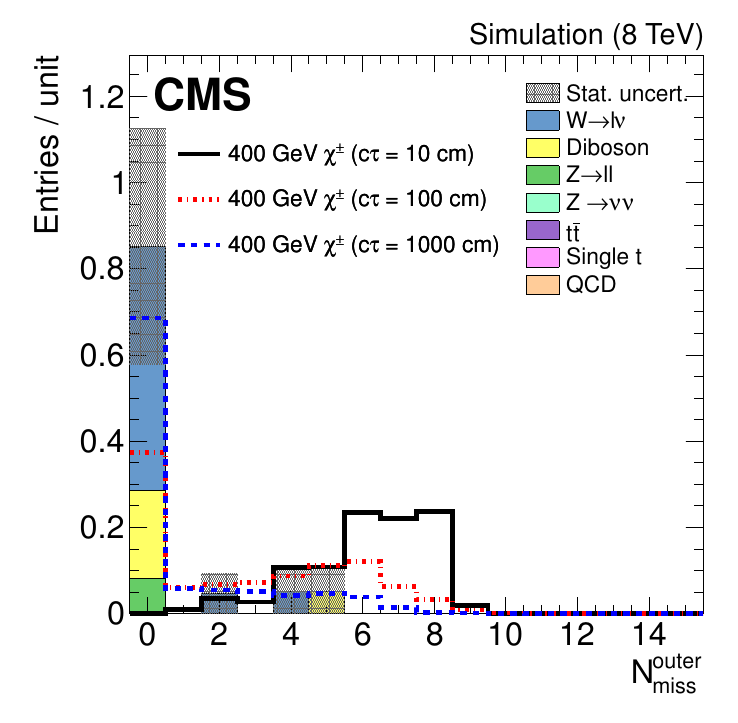}}
\end{minipage}
\hfill
\begin{minipage}{0.45\linewidth}
\centerline{\includegraphics[width=\linewidth]{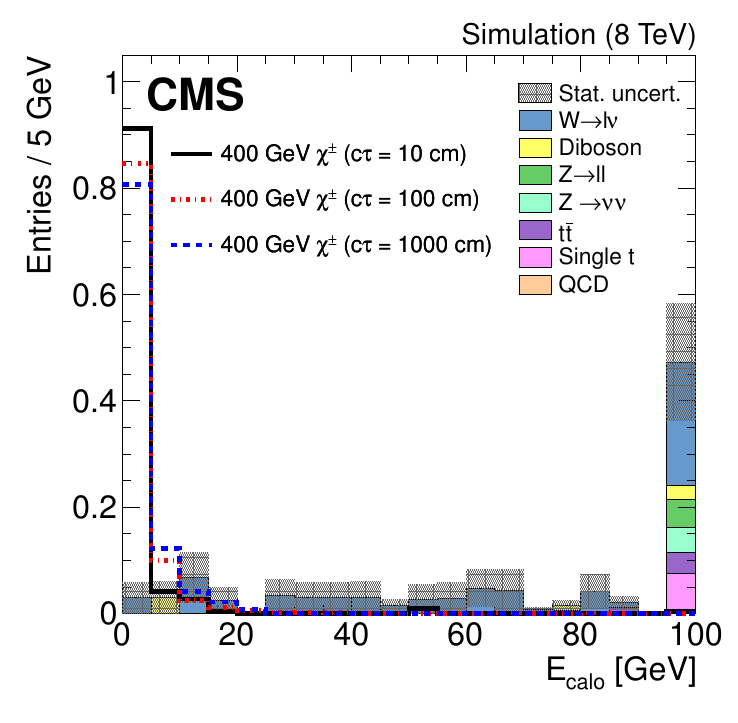}}
\end{minipage}
\caption{
The number of missing outer hits (left) and the associated calorimeter energy (right) of tracks in the search sample, before applying the requirement on the plotted quantity. The signal and the background sum distributions have both been normalized to unit area, and overflow entries are included in the last bin.
}
\label{fig:NmissoutEcalo}
\end{figure}

Candidate tracks are classified as disappearing if they meet two additional criteria.  First, they must have at least three missing outer hits.  These are tracker hits that are expected based on the track trajectory, but that are not recorded.   This requirement rejects most tracks from Standard Model particles, which typically produce hits in each tracker layer they intersect.
However, a track with missing outer hits may be produced by an electron that radiates a high-energy photon through bremsstrahlung or by a charged hadron that has a nuclear interaction with the detector material.  To avoid these sources of background, we additionally require that \Ecalo, the sum of the calorimeter energy in a cone of $\Delta R < 0.5$ around the candidate track, be less than 10 GeV.  Almost all of the signal passes this requirement, while the vast majority of the background is rejected.  
Simulated distributions of the number of missing outer hits and \Ecalo are shown in Figure~\ref{fig:NmissoutEcalo}, 

Tracks are sometimes reconstructed with missing outer hits when the track reconstruction software assigns the wrong set of hits.  The most common way this happens is that the track passes through the glue joint between two tracker sensor modules.  Since the glue joint is not an active sensor, no hit is recorded.  Sometimes this leads the tracking algorithm to select as the optimal track a trajectory with several missing outer hits.

\section{Backgrounds}
The main backgrounds confronted by this search arise from various failure modes of the reconstruction.  These are leptons that are unidentified, tracks with mismeasured momentum, or fake tracks.  

Electrons may produce a disappearing track if a large fraction of their energy is not reconstructed by the electromagnetic calorimeter.  To avoid this, we reject tracks in regions where the electromagnetic calorimeter has reduced efficiency, in the gap between the barrel and endcap, and dead or noisy crystals.  We also use a tag-and-probe study to find additional regions of electron inefficiency and veto tracks directed towards them.

Muons can produce disappearing tracks if they fail to be reconstructed as muons.  We therefore avoid any regions where the muon reconstruction efficiency is lower than average.  This occurs in two regions of pseudorapidity, the gap between the first and second wheels of the drift tube muon detectors, and a region between the inner and outer rings of the endcap muon stations.  We also veto tracks pointing toward muon chambers that are known to be faulty.  Muons may also fail to be reconstructed if they decay in flight or produce a secondary electromagnetic shower, but these are rare processes, occuring with a probability of $\sim10^{-5}$.  

Tau hadrons that produce a single track from a charged hadron may be classified as a disappearing track if the track momentum is mismeasured to a large degree.  In one simulated W$\rightarrow \tau (\pi \nu) \nu$ event, the reconstructed \pt of the pion is 75 GeV while its generated \pt is 15 GeV.  As a consequence, the pion track passes the \pt and \Ecalo requirements and meets the disappearing track criteria.  We require at least seven hits, which reduces the likelihood of mismeasuring the track momentum.  

Fake tracks are false trajectories that result from a pattern recognition failure of the track reconstruction algorithm.  They are mostly rejected by the track isolation and quality criteria, but in the rare case that they meet those criteria they closely mimic a track from signal.  

We estimate the background contribution for each of the background sources separately.  For the lepton backgrounds, we scale the yield in data control regions by an inefficiency factor obtained from simulation.  For the fake track background, we scale the number of events selected before applying any track criteria by the probability to reconstruct a fake track, which is obtained from a control sample.  

We validate the background estimates using control regions with inverted cuts on \Ecalo and on the number of missing outer hits, and find good agreement between the estimate and the observed yields.  
The estimated contribution from each background source is less than one event, and the total estimated background is $1.4 \pm 1.2$ events.

\section{Results}
In the data we observe two events with disappearing tracks, consistent with the background estimate.  The disappearing tracks in these two events are consistent with the estimated background distributions of \pt, \Ecalo, number of hits, and number of missing outer hits.

\begin{figure}[!h]
\centering
\begin{minipage}{0.45\linewidth}
\centerline{\includegraphics[width=\linewidth]{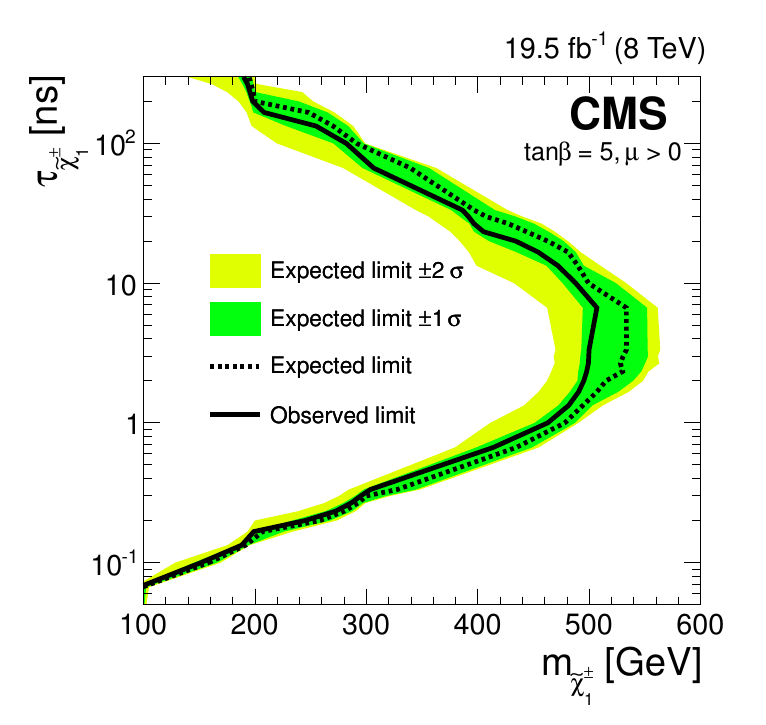}}
\end{minipage}
\hfill
\begin{minipage}{0.45\linewidth}
\centerline{\includegraphics[width=\linewidth]{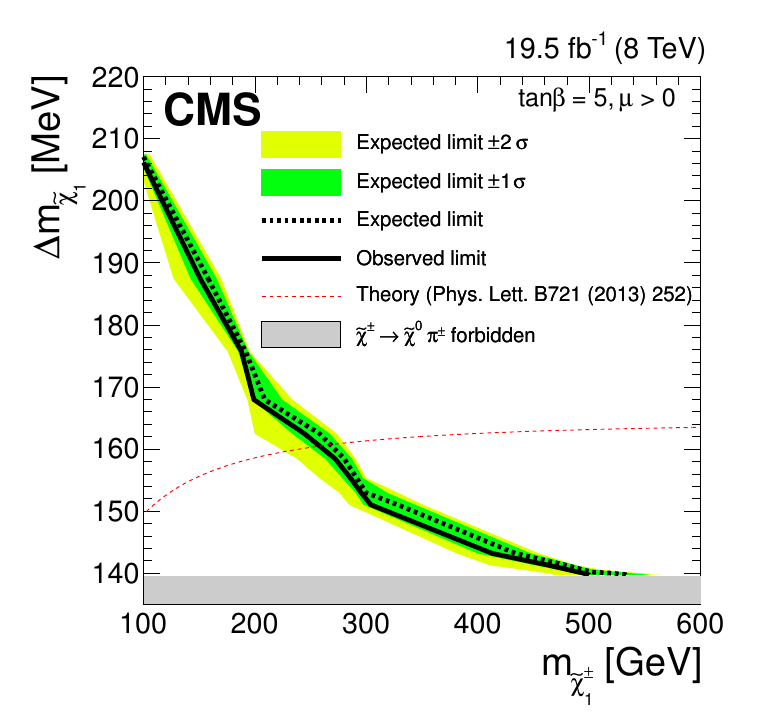}}
\end{minipage}
\caption{
The expected and observed constraints on the chargino mass and mean proper lifetime for generic electroweak chargino production (left) and on the chargino mass and the chargino-neutralino mass splitting in the AMSB model (right).  
}
\label{fig:Limits}
\end{figure}

\begin{figure}[!h]
\centering
\centerline{\includegraphics[width=0.6\linewidth]{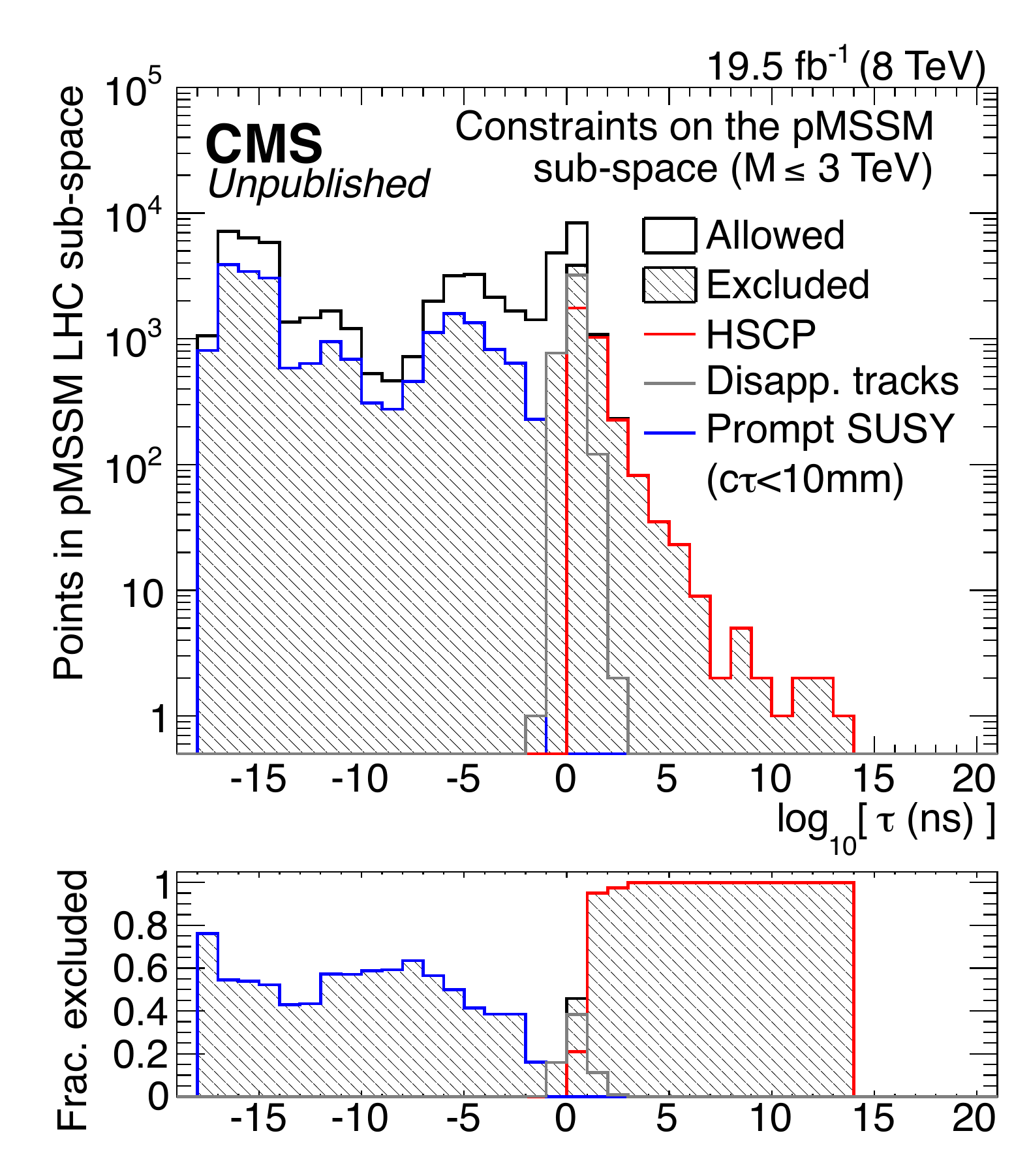}}
\caption{
Number of pMSSM points, in the sub-space covering sparticle masses up to about 3 TeV, that are excluded at a 95\% CL or allowed, as a function of the chargino lifetime.   
}
\label{fig:pMSSM}
\end{figure}

We place limits on generic electroweak chargino production and on chargino production in the AMSB framework, as shown in Figure~\ref{fig:Limits}.  The maximum sensitivity is to charginos with a lifetime of 7 ns, which we exclude for masses less than 505 GeV.  We also interpret the results in the framework of the phenomenological MSSM, and find that in a subspace of pMSSM points used to evaluate other CMS supersymmetry and exotic searches, the disappearing tracks search excludes model points corresponding to charginos with lifetime of about one nanosecond~\cite{EXO13006}, as shown in Figure~\ref{fig:pMSSM}.  

%
%

\section{Outlook}
Disappearing tracks could be a striking signature of physics beyond the Standard Model.  Using the Run 1 data, we have performed a search for disappearing tracks that sets limits on a benchmark model, AMSB.  This search probes a lifetime regime in between that targeted by searches for promptly decaying exotic particles and searches for heavy stable charged particles.  Since Run 1, we have implemented a new dedicated trigger which should improve the search sensitivity at $\sqrt{s} = 13$ TeV.  Now that we understand the several small background sources we are well-positioned in Run 2 to discover particles beyond the Standard Model that produce a disappearing track signature.

\Acknowledgments
I thank the other analysts who helped to produce the results presented in this talk.  I also thank all of our CMS colleagues, the CERN accelerator division for the excellent operation of the LHC, and the many funding agencies that support our work.

\end{document}